\documentclass[11pt,twoside]{article}
\usepackage{newpasp}

\newcommand{\HI}{\hbox{{\rm H}\kern 0.1em{\sc i}}}
\newcommand{\MgII}{\hbox{{\rm Mg}\kern 0.1em{\sc ii}}}
\newcommand{\FeII}{\hbox{{\rm Fe}\kern 0.1em{\sc ii}}}
\newcommand{\CIV}{\hbox{{\rm C}\kern 0.1em{\sc iv}}}
\newcommand{\NV}{\hbox{{\rm N}\kern 0.1em{\sc v}}}
\newcommand{\OVI}{\hbox{{\rm O}\kern 0.1em{\sc vi}}}

\newcommand{\Lya}{\hbox{{\rm Ly}\kern 0.1em$\alpha$}}
\newcommand{\CIVdblt}{{\rm C}\kern 0.1em{\sc iv}~$\lambda\lambda 1548, 1550$}
\newcommand{\MgIIdblt}{{\rm Mg}\kern 0.1em{\sc ii}~$\lambda\lambda 2796, 2803$}

\index{Charlton, J.C.}

\begin{document}

\title{From Globular Clusters to Tidal Dwarfs: Structure Formation in Tidal Tails}
\author{Jane Charlton$^{1}$, Karen Knierman$^{1,2}$, Sally Hunsberger$^{1}$,
Sarah Gallagher$^{1}$, Bradley Whitmore$^{3}$, Arunav Kundu$^{4}$,
\& John Hibbard$^{5}$}
\affil{$^{1}$The Pennsylvania State University, $^{2}$University of Arizona,
$^{3}$Space Telescope Science Institute, $^{4}$University of Virginia,
$^{5}$National Radio Astronomy Observatory} 


\begin{abstract}
Star clusters can be found in galaxy mergers, not only in central regions,
but also in the tidal debris.  In both the Eastern and Western tidal
tails of NGC 3256 there are dozens of young star clusters, confirmed by their 
blue colors and larger concentration index as compared to sources off 
of the tail.  Tidal tails of other galaxy pairs do not have such
widespread cluster formation, indicating environmental influences
on the process of star formation or the packaging of the stars.
\end{abstract}




\section{Introduction}
Extended structures resembling young dwarf galaxies are found in
tidal debris from galaxy interactions (Mirabel et~al. 1992;
Duc \& Mirabel 1994; Hunsberger, Charlton, \& Zaritsky 1996).
Star clusters form in
abundance in the central regions of interacting galaxy pairs
(Schweizer et~al. 1996; Miller et~al. 1997; Whitmore et~al. 1999;
Zepf et~al. 1999).
What physical conditions drive the formation of stars and determine
the nature of structure that forms in different environments?
Can star clusters also form in tidal debris, how widespread is
star formation in the debris, and how similar is it between different
tidal environments?

\section{HST WFPC2 Images of the Tidal Debris of NGC 3256}
The image in Figure 1 is a 1000 second exposure in the F555W
filter obtained on 1999 March 24.  A F814W image was also obtained.  
Point sources with $S/N > 3.0$ are indicated, those in the tail
with white circles and those out of the tail with white squares.
Although this field is crowded with foreground stars due to its 
low galactic latitude,  it is apparent that the numerous point 
sources are preferentially in the regions containing tidal debris.

\begin{figure}[th]
\plotfiddle{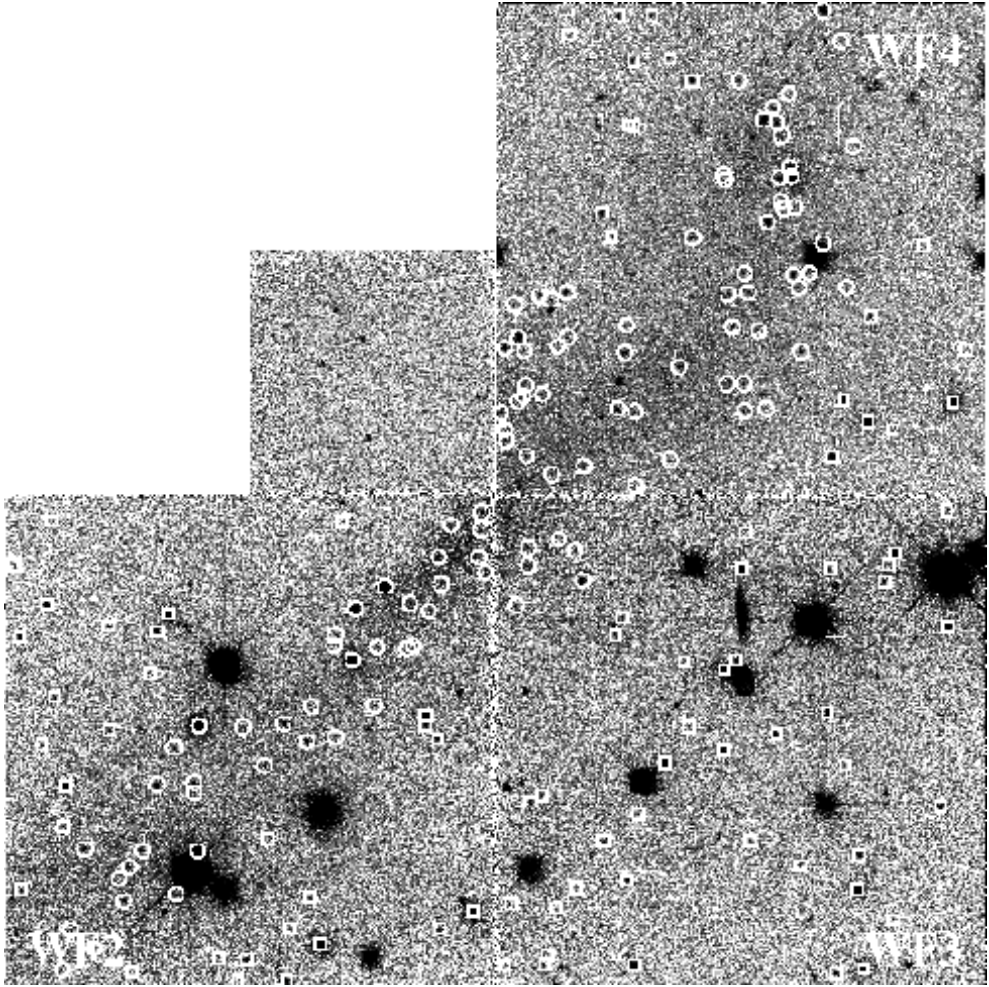}{4.4in}{0.}{76.}{76.}{-213}{-170}
\vglue 0.38in
\caption
{HST WFPC2 image of the Western tail of NGC 3256 taken with
the F555W ($V$--band) filter.  The sources within the tail are
identified with circles and those outside the tail with squares.
Sources were not detected at $S/N>3$ in the PC region.}
\end{figure}

\begin{figure}[htb]
\plotfiddle{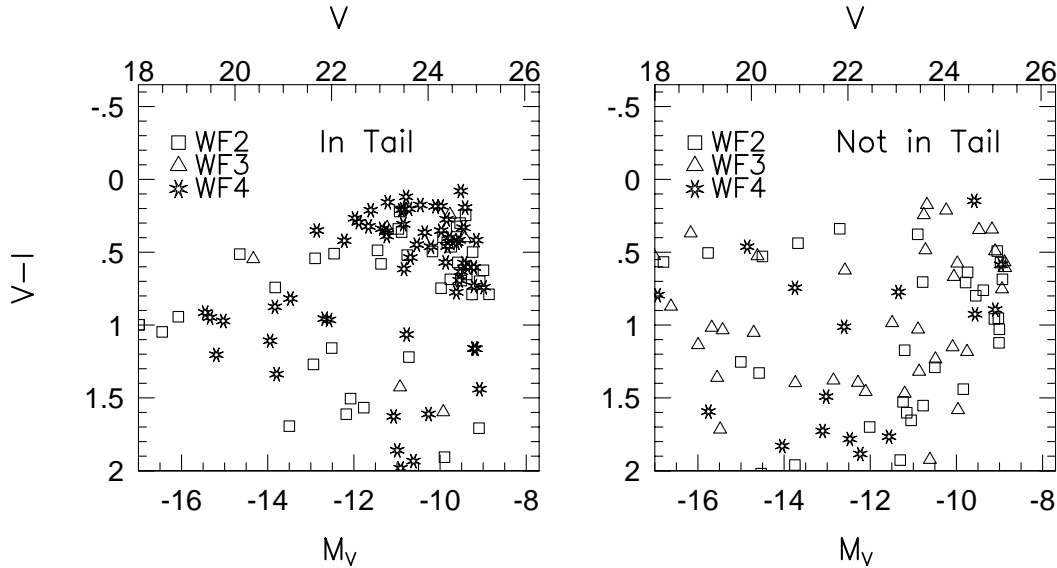}{3.in}{0.}{80.}{80.}{-254}{-370}
\vglue -0.4in
\caption
{Color Magnitude diagrams for sources in and out of the Western tail of 
NGC 3256.  The $V-I$ color is plotted vs. the absolute magnitude
on the lower horizontal scale and vs. the apparent $V$ magnitude
on the upper horizontal scale.  Sources are included if errors 
in $V$ are less than 0.25 mag, and if $-0.75 < V-I < 2.0$.  Sources
from the different WFPC2 chips are indicated by the symbols in
the legends.  Many of the ``in-tail'' sources occupy a region
in the relatively low luminosity, blue part of the diagram,
which is not heavily populated with ``out of tail'' sources.}
\end{figure}

\begin{figure}[ht]
\plotfiddle{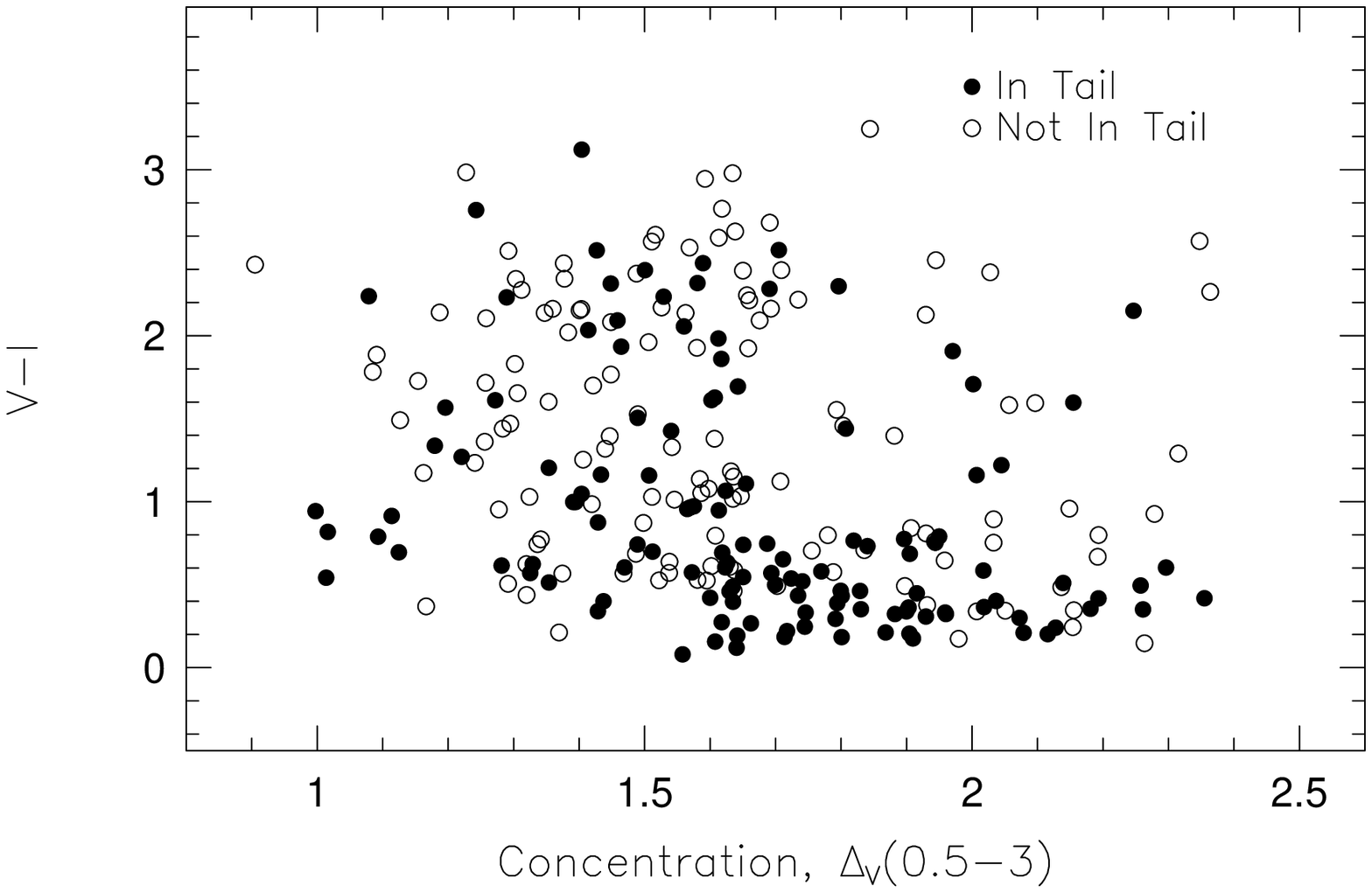}{4.0in}{0.}{75.}{75.}{-240.}{-240}
\vglue -0.8in
\caption
{ The concentration index is plotted vs. the $V-I$ color for
sources in (solid) and out (open) of the Western tail of 
NGC 3256.  The concentration
index, $\Delta_V (0.5 - 3)$, provides a rough measure of cluster size,
calculated from the difference in $V$ magnitude between aperture radii
of $0.5$ and $3.0$ pixels.
 }
\end{figure}

The colors, magnitudes, and sizes of these sources allow us to 
distinguish foreground and background contamination from star
clusters in the debris.  
Figure 2 illustrates that the brighter sources are relatively
young star clusters (i.e. hundreds of millions of years old), 
but some of the fainter ones could be individual stars.
The reddest sources are most certainly foreground stars with 
$-10 < M_V < -7.5$ apparent mostly from WF4.  There is an 
enhancement of point sources with $0 < V-I < 0.8$ and
$-12 < M_V < -9$.  The relatively large spread of the $V-I$ 
colors indicates either a range of ages, or non--uniform
extinction by dust.
The central region 
of NGC 3256 also has a large number of young clusters that
contribute 20\% of the total B--band luminosity of the galaxy
(Zepf et~al. 1999).

In Figure 3, for the Western tail, the $V-I$ color is plotted 
vs. the concentration index, defined as the difference between
$V$  magnitudes measured in a 0.5\arcsec and in a 3\arcsec aperture.
The solid circles represent sources in the tail regions while the
open circles represent sources in regions outside the tail. 
Clearly, the sources in the tail are on average larger, 
indicating that they are not point sources, and bluer.
Therefore, we confirm that there are many star clusters in
this tail.  In the Eastern tail of NGC 3256 we also find
a significant number of star clusters, but they are not
as abundant as in the Western tail.

\section{Environmental Influences on Cluster Formation in Tidal Debris}

We have also obtained HST/WFPC2 V and I band images of the tidal debris 
of three other mergers: NGC 4038/9, ``the Antennae'', NGC 7252,
``Atoms for Peace'', and NGC 3921.  
We detect several cluster candidates in NGC 7252 and NGC 4038/9,
and several super star clusters in the debris of NGC 7252 and
NGC 3921.
However, the debris of the remnant
NGC 3256 by far contains the largest number of massive star
clusters, both in the Eastern and in the Western tails.  
Apparently, the conditions in this remnant are
more conducive to formation of these clusters.

Clearly, not all tidal debris is equally conducive to the formation
of these clusters.  NGC 3256 is not distinguished from
the other pairs by age or by total {\HI} mass.
However, its two tidal tails are the only ones of the eight
we have studied that {\it do not} contain tidal dwarf galaxies.
NGC 7252, for example, contains in its Western tail a bright
dwarf with prominent patches of star formation, and in its
Eastern tail an extended low surface brightness dwarf.
In both dwarfs there are point sources with $V-I < 0.5$,
indicating that star formation continues well after the merger 
$\sim 750$ million years ago (Hibbard \& Mihos 1995).
Perhaps the formation of small stellar structures (star
clusters) and large stellar structures (tidal dwarfs) are
mutually exclusive.
Detailed comparisons of cluster positions to high resolution 21~cm 
maps of {\HI} content may also suggest factors that influence 
the formation and/or the packaging of stars.



\acknowledgements 
A detailed report of results on the tidal debris in all four
pairs, NGC 3256, NGC 7252, NGC 4038/9, and NGC 3921, has been
submitted to the Astronomical Journal (Knierman et~al. 2000).
This work was supported by NASA/STScI (grant GO-07466.01-96A).


\begin{references}
{\small

\reference Duc, P.-A., \& Mirabel, I. F. 1994, A\&A, 289, 83

\reference Hibbard, J. E., \& Mihos, J. C. 1995, AJ, 110, 140

\reference Hunsberger, S. D., Charlton, J. C., \& Zaritsky, D. 1996, ApJ, 462, 50

\reference Knierman, K. A., Hunsberger, S. D., Gallagher, S. C., Charlton, J. C., Whitmore, B. C., Kundu, A., Hibbard, J. E., \& Zaritsky, D. 2000, ApJ, submitted

\reference Miller, B. W., Whitmore, B. C., Schweizer, F., \& Fall, S. M. 1997, 
AJ, 114, 2381

\reference Mirabel, I. F., Dottori, H., \& Lutz, D. 1992, A\&A, 256, L19

\reference Schweizer, F., Miller, B. W.,  Whitmore, B. C.,\& Fall, S. M. 
1996, AJ, 112, 1839

\reference Whitmore, B. C., Zhang, Q.,  Leitherer, C., Fall, S. M.
 Schweizer, F., \& Miller, B. W. 1999, AJ, 118, 1551

\reference Zepf, S. et~al. 1999, AJ, 118, 752

}
\end{references}
\end{document}